\documentclass[twocolumn,pra,amsmath,amssymb,superscriptaddress]{revtex4-1}
\usepackage[utf8]{inputenc}
\usepackage{amsfonts,amsmath,amssymb}
\usepackage{graphicx}
\usepackage{bm} 
\usepackage{mathrsfs}
\usepackage{subfigure}
\usepackage{textcomp}
\usepackage{microtype}
\usepackage{hyperref}
\usepackage[flushleft]{threeparttable}
\usepackage[per-mode=symbol]{siunitx}
\usepackage[version=3]{mhchem}

\usepackage{booktabs}
\DeclareSIUnit\intensity{\watt\per\centi\meter\squared}
\DeclareSIUnit\fieldstrength{\volt\per\centi\meter}
\usepackage{xspace}

\DeclareUnicodeCharacter{2009}{\,}

\newcommand{\cost}{\ensuremath{\langle\cos^2\theta_\text{2D}\rangle}}

\newcommand{\singlet}{$1^1\Sigma_g^+$\xspace}
\newcommand{\triplet}{$1^3\Sigma_u^+$\xspace}

\newcommand{\req}{$R_\text{eq}$\xspace}
\newcommand{\bgas}{$B_\text{gas}$\xspace}
\newcommand{\bhe}{$B_\text{He}$\xspace}
\newcommand{\btd}{$B_{\text{2D}}$\xspace}
\newcommand{\deltaB}{$\Delta B_\text{He}$\xspace}

\newcommand{\vhe}{$V_\text{He}$($\Theta$)\xspace}
\newcommand{\ptheta}{$P(\Theta)$\xspace}

\newlength{\figwidth}
\newlength{\figwidthwide}
\let\orgautoref\autoref
\providecommand{\Autoref}{%
  \def\equationautorefname{Equation}%
  \def\figureautorefname{Figure}%
  \def\subfigureautorefname{Figure}%
  \def\tableautorefname{Table}
  \def\sectionautorefname{Section}%
  \orgautoref}
\renewcommand{\autoref}{%
  \def\equationautorefname{Eq.}%
  \def\figureautorefname{Fig.}%
  \def\subfigureautorefname{Fig.}%
  \def\sectionautorefname{Sec.}%
  \orgautoref}

\usepackage{color}
\usepackage[normalem]{ulem}
\definecolor{darkgreen}{rgb}{0.0,0.7,0.0}

\newcommand{\ket}[1]{\, | #1 \rangle}

\begin{document}

\title{Nonadiabatic laser-induced alignment dynamics of alkali dimers on the surface of a helium droplet} 

\author{Henrik H. Kristensen}
\affiliation{Department of Chemistry, Aarhus University, Langelandsgade 140, DK-8000 Aarhus C, Denmark}
\author{Lorenz Kranabetter}
\affiliation{Department of Chemistry, Aarhus University, Langelandsgade 140, DK-8000 Aarhus C, Denmark}
\author{Areg Ghazaryan}
\affiliation{Institute of Science and Technology Austria, Am Campus 1, 3400 Klosterneuburg, Austria}
\author{Constant A. Schouder}
\affiliation{Department of Chemistry, Aarhus University, Langelandsgade 140, DK-8000 Aarhus C, Denmark}
\affiliation{ISMO, CNRS, Universit\'{e} Paris‐Saclay, 91400 Orsay, France}
\author{Emil Hansen}
\affiliation{Department of Physics and Astronomy, Aarhus University, Ny Munkegade 120, 8000 Aarhus C, Denmark}
\author{Frank Jensen}
\affiliation{Department of Chemistry, Aarhus University, Langelandsgade 140, DK-8000 Aarhus C, Denmark}
\author{Robert E. Zillich}
\affiliation{Institute for Theoretical Physics, Johannes Kepler Universität Linz, Altenbergerstra\ss{e} 69, A-4040 Linz, Austria}

\author{Mikhail Lemeshko}
\affiliation{Institute of Science and Technology Austria, Am Campus 1, 3400 Klosterneuburg, Austria}

\author{Henrik Stapelfeldt}
\email[]{henriks@chem.au.dk}
\affiliation{Department of Chemistry, Aarhus University, Langelandsgade 140, DK-8000 Aarhus C, Denmark}

\date{\today}

\begin{abstract}

Alkali dimers, \ce{Ak2}, located on the surface of a helium nanodroplet, are set into rotation through the polarizability interaction with a nonresonant 1-ps-long laser pulse. The time-dependent degree of alignment is recorded using femtosecond-probe-pulse-induced Coulomb explosion into a pair of \ce{Ak+} fragment ions. The results, obtained for \ce{Na2}, \ce{K2}, and \ce{Rb2} in both the ground state \singlet and the lowest-lying triplet state \triplet, exhibit distinct, periodic revivals with a gradually decreasing amplitude. The dynamics differ from that expected for dimers had they behaved as free rotors. Numerically, we solve the time-dependent rotational Schrödinger equation, including an effective mean-field potential to describe the interaction between the dimer and the droplet. The experimental and simulated alignment dynamics agree well and their comparison enables us to determine the effective rotational constants of the alkali dimers with the exception of \ce{Rb2}(\triplet) that only exhibits a prompt alignment peak but no subsequent revivals. For \ce{Na2}(\triplet), \ce{K2}(\singlet), \ce{K2}(\triplet) and \ce{Rb2}(\singlet), the alignment dynamics are well-described by a 2D rotor model. We ascribe this to a significant confinement of the internuclear axis of these dimers, induced by the orientation-dependent droplet--dimer interaction, to the tangential plane of their residence point on the droplet.

\end{abstract}

\maketitle 

\section{Introduction}\label{sec:intro}

Laser-induced alignment, the confinement of molecular axes to fixed-in-space axes through the use of moderately intense laser pulses, has been intensively explored for almost 30 years~\cite{stapelfeldt_colloquium:_2003,seideman_nonadiabatic_2005-1,ohshima_coherent_2010,fleischer_molecular_2012,koch_quantum_2019}. A particular emphasis has been on alignment in the nonadiabatic regime where a femtosecond or picosecond laser pulse creates a rotational wave packet in each molecule. This leads to alignment dynamics characterized by periodically spaced, narrow time intervals, termed revivals, in which the degree of alignment reaches local maxima and minima. The majority of the many works reported explored isolated gas-phase molecules but interest also turned towards molecules in dissipative environments~\cite{ramakrishna_intense_2005}. One direction of studies investigated molecules in dense gasses~\cite{vieillard_field-free_2008,owschimikow_cross_2010} allowing, e.g., non-Markovian collisional dynamics and related phenomena to be studied~\cite{zhang_rotational_2019,ma_observing_2019,bournazel_non-markovian_2023}.

Over the last decade, it has been demonstrated that laser-induced alignment methods can also be applied to molecules embedded in liquid helium, mainly in the shape of nanodroplets~\cite{pentlehner_impulsive_2013,schmidt_rotation_2015,pickering_alignment_2018,vindel-zandbergen_impulsive_2018,chatterley_long-lasting_2019,cherepanov_excited_2021,qiang_femtosecond_2022,liu_theory_2023} and very recently also in bulk samples~\cite{milner_coherent_2024}. In the nonadiabatic regime, studies showed that for low laser intensities, the observed alignment dynamics closely resembled that determined from solution of the rotational time-dependent Schr\"odinger equation for isolated molecules using the in-helium effective rotational constants and accounting for inhomogeneous broadening~\cite{chatterley_rotational_2020}. The explanation was that the rotational energy of the molecules obtained from the laser interaction remained low enough that coupling to rotons and phonons of the He droplet is very weak. As such, the He-dressed molecules, i.e., the molecules along with their solvation shell of helium atoms, could rotate almost freely. At higher intensities, the alignment dynamics differs significantly from the gas-phase case leading to, e.g., transient decoupling of the molecules from its solvation shell~\cite{shepperson_laser-induced_2017}.

Most molecules are located in the interior of helium nanodroplets but a few species, notably dimers of alkali atoms, are known to reside on the surface~\cite{ancilotto_sodium_1995,stienkemeier_electronic_2001, barranco_helium_2006}. This offers an opportunity for exploring laser-induced alignment of molecules in a novel regime namely on a surface. In 2023, we reported nonadiabatic alignment dynamics of \ce{Na} dimers in the \triplet state on the surface of helium nanodroplets~\cite{kranabetter_nonadiabatic_2023}. The time-dependent degree of alignment exhibited a revival pattern that differed qualitatively from that expected for freely rotating molecules. The difference was shown to originate from the surface potential of the droplet, which confines the orientation of the dimers to the tangential plane of the surface at the point where the dimer is bound. The confinement is sufficiently strong that the observed alignment dynamics is well-described by that of a 2D quantum rotor. Furthermore, by comparing the observed alignment dynamics to results from quantum simulations, it became possible, for the first time, to determine the effective rotational constant, \bhe. Its value is $\sim$ 1.3 times smaller than the value of the gas-phase rotational constant. This reduction factor is less than in the case of diatomic and linear triatomic molecules inside He droplets~\cite{choi_infrared_2006}. The difference was ascribed to the fact that \ce{Na2} interacts more weakly with the He droplet compared to molecules localized in the droplet interior. Here, we present a more comprehensive account of laser-induced alignment dynamics of alkali dimers on helium droplet surfaces. This is done through studies of \ce{Na2}, \ce{K2}, and \ce{Rb2} in both the \singlet and the \triplet state. These systems represent different polarizabilities, different rotational constants and different droplet-dimer interactions, parameters that all are expected to influence the alignment dynamics.

\section{Experimental setup}\label{sec:setup}

The apparatus used for the experiments has been described previously~\cite{kristensen_laser-induced_2023}. Thus, we only provide the essential details here. A beam of He droplets (estimated mean size of 15000 He atoms) from a continuous source is passed through a pickup cell containing a gas of either \ce{Na}, \ce{K}, or \ce{Rb}. By regulating the temperature of the pickup cell, the vapor pressure is adjusted to a value where some of the droplets pick up two \ce{Ak} atoms, leading to the formation of an alkali dimer in the \singlet state or the \triplet state~\cite{stienkemeier_laser_1995, higgins_helium_1998, aubock_triplet_2007}. The now doped droplet beam continues forward into a velocity map imaging (VMI) spectrometer, inside of which it is crossed by two focused, linearly polarized laser beams. The pulses in the first beam are used to induce alignment of the alkali dimers. They have a duration, $\tau_\text{a}$, in the range 0.6--1.0~ps (FWHM), the exact values are given on \autoref{fig:all_traces}, and a central wavelength of 1300~nm. At this wavelength, the photon energy falls below the energy needed to cause single-photon electronic transitions from either the \singlet or the \triplet state~\cite{magnier_potential_1993, magnier_theoretical_2004, allouche_transition_2012}, i.e., linear absorption is eliminated. The pulses are focused to a spot size of $\omega_0$~=~85~$\mu\text{m}$ and the peak intensities of the pulses, $I_\text{a}$, given on \autoref{fig:all_traces}, are chosen sufficiently high to induce significant alignment of the dimers yet sufficiently low that multiphoton ionization is avoided. The pulses in the second beam ($\lambda$~=~800~nm [400~nm for \ce{Rb2} and for \ce{K2} (\singlet)], $\tau_{\text{FWHM}}$~=~50~fs), sent at a delay $t$ after the alignment pulses, are used to doubly ionize the dimers through multiphoton absorption, \autoref{fig:images_potentials}(c), which leads to Coulomb explosion into pairs of \ce{Ak+} ions~\cite{kristensen_quantum-state-sensitive_2022, kristensen_laser-induced_2023}. As described below, detection of the emission direction of these ions enables the determination of the degree of alignment at time $t$. In practice, the VMI spectrometer projects the \ce{Ak+} ions onto a 2D imaging detector backed by a CCD camera. For \ce{Na2}, \ce{K2}, and \ce{Rb2} the detector is gated such that only \ce{^{23}Na^+}, \ce{^{39}K^+}, and \ce{^{85}Rb^+} ions, respectively, are recorded. The camera acquires frames containing the ion hits from 10 consecutive sets of alignment and probe laser pulses. For the probe pulses, intensities ranging from 5$\times$10$^{12}$ to 9.3$\times$10$^{13}$~W/cm$^2$ are used with a focused spot size of $\omega_0$~=~40~$\mu\text{m}$.

\section{Experimental results}\label{sec:Exp_results}

\subsection{Ion images}

\begin{figure}
\includegraphics[width = 8.6cm]{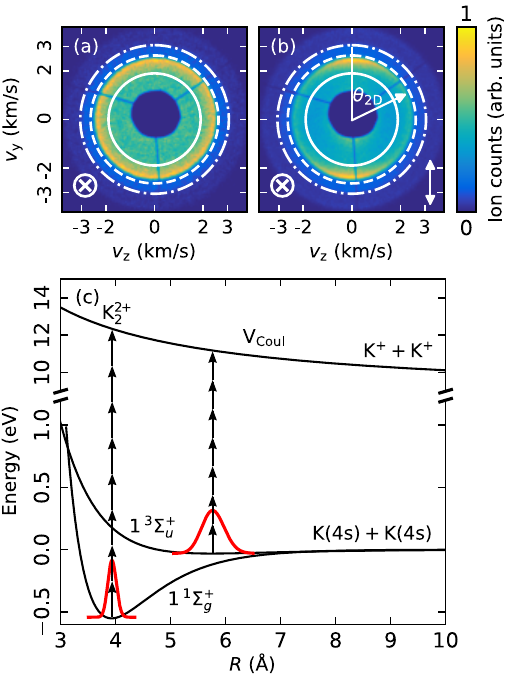}
\caption{(a) 2D-velocity ion image for \ce{K+} recorded with the probe pulse only. (b) \ce{K+} ion image recorded with alignment and probe pulse at $t$~=~10~ps. The polarization directions of the probe pulses ($\otimes$) and the alignment pulses ($\updownarrow$) are illustrated at the bottom corners of the images. (c) Potential energy curves for the \singlet~\cite{magnier_theoretical_2004} and \triplet~\cite{bauer_accurate_2019} states of \ce{K2} and the Coulomb potential. The red contours show the vibrational ground state wave functions for the \singlet and \triplet potentials. The vertical black arrows represent the laser photons (not to scale) required to doubly ionize the dimers, which initiates the Coulomb explosion.}
\label{fig:images_potentials}
\end{figure}

\Autoref{fig:images_potentials}(a) shows a 2D-velocity image of \ce{K+} ions recorded using the probe pulse only. Meanwhile, \autoref{fig:images_potentials}(b) shows an image of \ce{K+} ions obtained when using both the alignment and the probe pulse. The images lack signal in the center and in three radial stripes, due to a centrally positioned metal disk and its three mounting rods~\cite{schouder_laser-induced_2020, chatterley_laser-induced_2020}. The main purpose of this disk is to block out \ce{Ak+} ions from ionization of effusive \ce{Ak} atoms that entered the VMI spectrometer. Outside the center, a few distinct channels are visible in the images, marked by annotated white circles. To understand the origin of these channels it is useful to consult the potential energy diagram for \ce{K2}, displayed in~\autoref{fig:images_potentials}(c).

The probe pulse doubly ionizes \ce{K2}, corresponding to projection of the initial wave function for the $v$~=~0 vibrational level \footnote{This is the only vibrational level populated at the 0.4 K temperature of the droplet} in either the \singlet state or the \triplet state onto the \ce{K2^{2+}} potential curve. For the internuclear distances considered, this curve is to a good approximation given by a Coulomb potential~\cite{kristensen_laser-induced_2023}. Subsequently, the molecular dication breaks apart into two \ce{K+} ions, the process termed Coulomb explosion. The final kinetic energy of each \ce{K+} ion equals half of the initial Coulomb energy $V_{\text{Coul}}(R_\text{eq}) = 14.4$ eV/$R_\text{eq}$ [\AA], where $R_\text{eq}$ is the equilibrium bond distance~\footnote{This holds for the \ce{^{39}K2} isotopologue (87.0 \% abundance). For the \ce{^{39}K^{41}K} isotopologue (12.6 \% abundance), the \ce{^{39}K+} ion gets 51.3 \% of the Coulomb energy.}. For the \singlet state $R_\text{eq}$~=~3.92~\AA{} and for the \triplet state $R_\text{eq}$~=~5.73~\AA~\cite{deiglmayr_calculations_2008, bauer_accurate_2019}.  Consequently, \ce{K+} ions originating from dimers in the \singlet (\triplet) state end with $E_\text{kin}$ = 1.84 eV (1.26 eV). This difference in the final $E_\text{kin}$  allow us to identify the ions in the region between the solid and dashed rings (between the dashed and dot-dashed rings) in~\autoref{fig:images_potentials}(a) and (b) as those from Coulomb exploded \triplet (\singlet) state dimers. For \ce{Na2} and \ce{Rb2}, it is possible to make a similar distinction between the two quantum states in the 2D velocity images of the \ce{Ak+} ion, see Refs.~\cite{kristensen_quantum-state-sensitive_2022, kristensen_laser-induced_2023} for details. Therefore, through analysis of the ion hits in the two channels, we are able to characterize the alignment dynamics of the three different alkali dimers studied in both the \singlet and the \triplet state.

When only the probe pulse, polarized perpendicular to the detector plane, is present, the velocity image, \autoref{fig:images_potentials}(a), is circularly symmetric. This is the result expected for randomly oriented molecules. When the alignment pulse is applied prior to the probe pulse the circular symmetry is broken.  For the image displayed in \autoref{fig:images_potentials}(b), where the probe pulse is sent at $t$~=~10~ps, the \ce{K+} ions in both the \singlet and \triplet channel exhibit a clear anisotropy centered along the polarization axis (vertical) of the alignment pulse. In line with many previous works~\cite{pentlehner_impulsive_2013,pentlehner_laser-induced_2013}, we interpret this observation as alignment of the dimers along the alignment pulse polarization direction.

\subsection{Alignment dynamics}

\begin{figure*}
\includegraphics[width=17.8 cm]{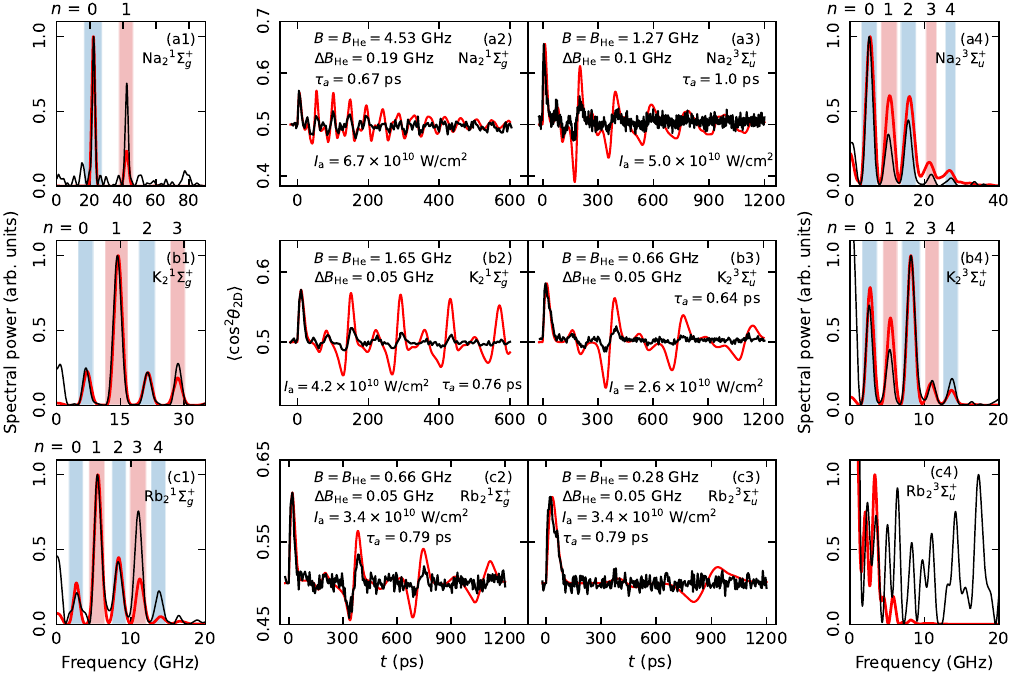}
\caption{Time-dependent degree of alignment for (a2)--(c2) \ce{Ak2} in the \singlet state and (a3)--(c3) \ce{Ak2} in the \triplet state. The black curves show the experimental data and the red curves show the simulated results obtained for the 3D model + \vhe, \autoref{eqn:hamil}, including inhomogeneous broadening. (a1)--(c1) and (a4)--(c4) show the power spectra for the corresponding \cost{} traces in (a2)--(c2) and (a3)--(c3), respectively. The spectral peaks are assigned numbers $n$, given above each panel, in order and highlighted with colored bands (blue: even $n$, red: odd $n$).}
\label{fig:all_traces}
\end{figure*}

The alignment dynamics were measured for \ce{Na2}, \ce{K2}, and \ce{Rb2} in both the \singlet state and the \triplet state by recording ion images for various delays $t$ between the alignment and probe pulse. The delay was scanned from $t = -20$~ps to $t = 600$~ps for \ce{Na2} and \ce{K2} in the \singlet state, while a range from $t = -20$~ps to $t = 1200$~ps was used for \ce{Rb2} in the \singlet state and for the measurements on the \triplet states of all the dimers. For all measurements the step size was kept between 1~ps and 5~ps. To quantify the degree of alignment, we determine the widely used metric \cost{} for the ions in the \singlet and \triplet radial ranges individually for each image. Here, $\theta_{\text{2D}}$ is the angle in the detector plane between the ion hit and the alignment pulse polarization direction, see~\autoref{fig:images_potentials}(b). \Autoref{fig:all_traces}(a2)--(c2) and (a3)--(c3) show \cost$(t)$ for the \singlet state and \triplet state dimers, respectively.

\Autoref{fig:all_traces}(a3) shows the results for \ce{Na2} in the \triplet state that were also presented in Ref.~\cite{kranabetter_nonadiabatic_2023}. Immediately after the alignment pulse is over, \cost{} rises from 0.5, the value characterizing randomly oriented molecules, to the global maximum at $t$~=~6~ps, where \cost{}~=~0.66. At later times, \cost{} exhibits an oscillatory structure with an amplitude that gradually diminishes. After $\sim$~1000~ps, \cost{} is back to a value of 0.5 and the oscillations are hardly visible any longer. The result for \ce{K2}(\triplet state), \autoref{fig:all_traces}(b3), is qualitatively similar to that for \ce{Na2}(\triplet). Again, \cost{} reaches a global maximum at early times ($t$~=~15~ps: \cost{}~=~0.58). The ensuing alignment dynamics is slower, notably the periodicity of the oscillations is larger, which appears reasonable since the moment of inertia of an isolated \ce{K2}(\triplet state) dimer is about twice as large as that of an isolated \ce{Na2}(\triplet state) dimer~\cite{ladjimi_diatomic_2024}. In the case of \ce{Rb2}(\triplet), \autoref{fig:all_traces}(c3), there is also an initial increase in \cost{} to a maximum of 0.60 at $t$~=~25~ps. However, no oscillatory structure is observed at later times.

Turning to the \singlet state results, shown  in \autoref{fig:all_traces}(a2)--(c2), we note that each of these \cost$(t)$ traces also exhibit an initial alignment peak, followed by oscillations with recurring revivals that gradually decrease in amplitude. It is striking that several revivals are present for \ce{Rb2} in the \singlet state, in contrast to \ce{Rb2} in the \triplet state. A common trait for all six alignment traces measured is that they differ qualitatively from those observed for linear molecules in gas phase, see a few selected examples in Refs.~\cite{dooley_direct_2003, ghafur_impulsive_2009, wu_nonadiabatic_2011, chatterley_laser-induced_2020}, and embedded inside helium droplets~\cite{chatterley_rotational_2020}.

\begin{figure}
\includegraphics[width=8.6cm]{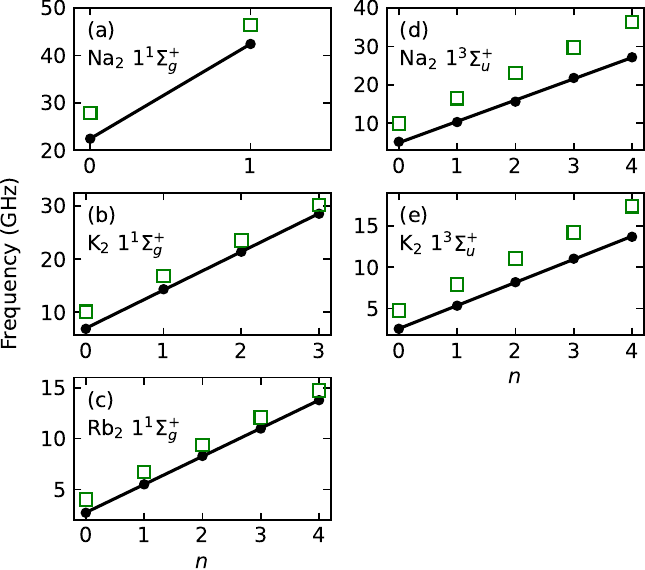}
\caption{Central positions, marked with black dots, of the frequency peaks in \autoref{fig:all_traces}(a1)--(c1) and (a4)--(b4), plotted as a function of the assigned $n$. Linear fits to the data points are shown with black lines. The green squares show the expected gas-phase frequencies $(4J+6)B_\text{gas}$ calculated for $J = n$.}
\label{fig:FFT}
\end{figure}

To analyse the results, we Fourier transform the alignment traces, thereby obtaining the spectral content. The black curves in \autoref{fig:all_traces}(a1)--(c1) and (a4)--(c4) show the spectra corresponding to the \cost$(t)$ curves in \autoref{fig:all_traces}(a2)--(c2) and (a3)--(c3), respectively. Distinct peaks are observed in all the spectra, except for the case of \ce{Rb2}(\triplet), see \autoref{fig:all_traces}(c4), where only noise is present. We label the peaks with integer values $n$ in order of increasing frequency, annotated above the panels in \autoref{fig:all_traces}(a1)--(c1) and (a4)--(b4). The central positions of the spectral peaks plotted as a function of the assigned $n$ (black dots in \autoref{fig:FFT}) fall very closely on a straight line. This linear relationship is corroborated by linear fits, shown with black lines in \autoref{fig:FFT}, to the data points. For rigid, linear, gas-phase molecules the peak positions are given by $(4J+6)B_\text{gas}$~\cite{chatterley_rotational_2020}. Using the literature values for \bgas listed in \autoref{tab:B-values}, we calculate the peak positions for each dimer with $J = n$. The results are shown with green squares in \autoref{fig:FFT}. For \ce{Na2} and \ce{K2} in the \triplet state, \autoref{fig:FFT}(d) and (e), the experimental peak positions deviate strongly from those expected for the isolated rotor. Similarly, there is a pronounced discrepancy between the gas-phase model and the observations for the \singlet state measurements of all three dimers, see \autoref{fig:FFT}(a)--(c). Thus, the spectral analysis supports that the observed alignment dynamics are not just that of linear molecules rotating in the gas phase.

\section{Modeling the alignment dynamics}\label{sec:theory}

To model the alignment dynamics of the dimers, we solve the time-dependent rotational Schrödinger equation for the Hamiltonian:
\begin{align}
\hat{H} =  \frac{B\hat{J}^2}{\hbar^2} - \frac{1}{4}E(t)^2 \Delta\alpha \cos^2\theta + V_\text{He}(\Theta).\label{eqn:hamil}
\end{align}

Here, the first two terms are the same as those used when modelling laser-induced alignment of linear molecules in gas phase~\cite{friedrich_alignment_1995, fleischer_molecular_2012}. $B$ is the rotational constant of the molecule, $\hat{J}^2$ is the squared rotational angular momentum operator, and $\Delta\alpha$ is the polarizability anisotropy of the molecule. The electric field envelope of the alignment pulse is denoted by $E(t)$ and the angle between the internuclear axis and the polarization axis of the linearly polarized alignment pulse is denoted by $\theta$. The third term, \vhe, is an effective mean-field potential used to model the interaction between the alkali dimer and the He droplet with $\Theta$ denoting the angle between the dimer internuclear axis and the droplet surface normal.

To obtain \vhe we first calculated the \ce{Ak2}-He potentials~\footnote{See Supplemental Material at [URL will be inserted by publisher] for the \ce{Ak2}-He potentials} at the CCSD(T) level with the valence and outermost core orbitals included in the correlation treatment and with basis set extrapolation~\cite{halkier_basis-set_1999} and using half of the counterpoise correction~\cite{burns_comparing_2014,van_duijneveldt_state_1994}. For Na, the aug-cc-pVXZ (X = D,T,Q,5) basis sets were used~\cite{prascher_gaussian_2011}. For K, the aug-cc-pVXZ-x2c (X = D,T,Q)~\cite{hill_gaussian_2017} were used with a recontraction using Hartree-Fock atomic coefficients, and for Rb, the Sapporo nZP (Z = D, T, Q) were used~\cite{noro_segmented_2012}.

\begin{figure}
\includegraphics[width = 8.6cm]{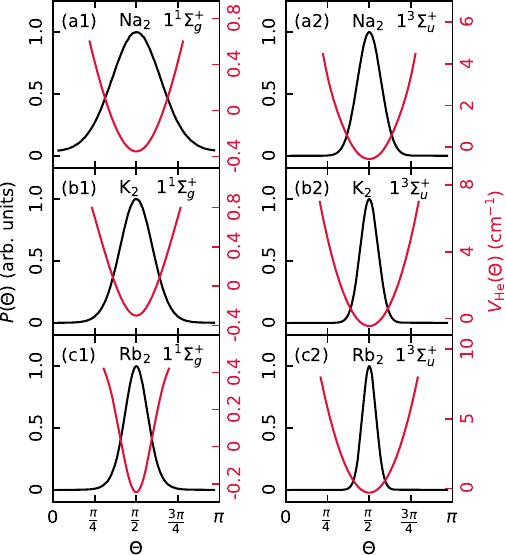}
\caption{Effective potentials \vhe (shown in red) between the helium droplet surface and the alkali dimers in (a1)--(c1) the \singlet state and (a2)--(c2) the \triplet state. The angular distribution function $P(\Theta)$ for each dimer is shown in black. $\Theta$ denotes the angle between the internuclear axis of the dimer and the surface normal.}
\label{fig:confinement}
\end{figure}

Next, the \ce{Ak2}--\ce{He} potentials were used to calculate \vhe for the three different alkali dimers in both the \singlet state and the \triplet state. We follow the procedure described in detail in Ref.~\cite{guillon_theoretical_2011} and the supplement of Ref.~\cite{kranabetter_nonadiabatic_2023}: we approximate the curved surface of large \ce{He} droplets with the flat surface of a \ce{He} film on which an \ce{Ak2} molecule is adsorbed. From path integral Monte Carlo (PIMC) simulations we obtain the orientational distribution $P(\Theta)$, i.e., the probability density that the dimer axis has an angle $\Theta$ from the surface normal. The temperature is chosen low enough ($T=0.31$~K, which is slightly lower than $T=0.37$~K assumed in experiments) that the \ce{Ak2} essentially occupies its ground state. On the other hand, we use the model Hamiltonian, \autoref{eqn:hamil}, and equate its ground state with the PIMC result, $\Phi_0(\Theta)=\sqrt{P(\Theta)}$. We can thus solve the effective Schr\"odinger equation for $V_\text{He}$, and obtain $V_\text{He}(\Theta) = -\frac{1}{\Phi_0(\Theta)}\frac{B\hat J^2}{\hbar^2}\Phi_0(\Theta)$, apart from a constant that is irrelevant for the dynamics. For \ce{Rb2} in the \triplet state, \ptheta and \vhe were taken from Ref.~\cite{guillon_theoretical_2011}. \Autoref{fig:confinement} shows \vhe and \ptheta for \ce{Na2}, \ce{K2}, and \ce{Rb2}.  For all three dimers in both the \singlet and the \triplet state, \ptheta peaks at $\Theta = \pi/2$, i.e, the dimers are preferentially oriented parallel to the droplet surface. We note that for each dimer species, the \ptheta distribution for the \singlet state is wider than for the \triplet state, and that for both the singlet and the triplet state, the angular distributions narrows as the mass of the dimer increases.

The Schrödinger equation with the Hamiltonian from~\autoref{eqn:hamil}, is solved using the effective potentials \vhe and gas-phase values for $\Delta\alpha$ taken from the literature~\cite{deiglmayr_calculations_2008,ladjimi_diatomic_2024}. We use effective rotational constants \bhe instead of the gas-phase rotational constants and include inhomogeneous broadening of the rotational levels as done in Ref.~\cite{kranabetter_nonadiabatic_2023} and in the modeling of alignment of molecules inside helium droplets~\cite{chatterley_rotational_2020}. The inhomogeneous broadening accounts for the different local environments experienced by the dimers due to the variation of droplet sizes present in the experiment~\cite{lehmann_potential_1999,lehmann_lorentzian_2007,zillich_lineshape_2008}. We implement the inhomogeneous broadening as a Gaussian distribution of \bhe with a FWHM of \deltaB. In the numerical calculations, \bhe and \deltaB are treated as free parameters and their values are varied to optimize the agreement between the calculated and measured \cost{}. This is the same strategy as in rotational coherence spectroscopy (RCS), where the comparison of time-dependent alignment-sensitive experimental signals to results from numerical calculations are used to determine rotational constants~\cite{felker_rotational_1992,riehn_high-resolution_2002}.

We describe the initial rotational population by a Boltzmann distribution at the 0.37~K temperature expected for the alkali dimers on the helium droplets~\cite{aubock_triplet_2007}, and include weighting of the even and odd $J$ states according to nuclear spin statistics~\footnote{For \ce{Na2} and \ce{^{39}K2}:  \singlet: odd/even: 5/3; \triplet: odd/even: 3/5. For \ce{^{85}Rb2}: \singlet: odd/even: 7/5; \triplet: odd/even: 5/7.}. For the experiments with \ce{K2} and \ce{Rb2}, the presence of multiple isotopologues were also taken into account. Due to the isotope abundances~\cite{berglund_isotopic_2011}, 93.3\% of the \ce{^39K+} ions recorded originate from \ce{^39K_2}, while 6.7\% come from \ce{^39K^41K}. Similarly, 72.2\% of the \ce{^85Rb^+} ions measured stem from \ce{^85Rb_2} and 27.8\% from \ce{^85Rb^87Rb}. The isotope abundances were included as weights in the calculation. In the \ce{Na} case there is only the \ce{^23Na} isotope. We did not include focal volume averaging because the focal spot size of the alignment beam is more than twice as big as that of the probe beam~\footnote{In Ref.~\cite{kranabetter_nonadiabatic_2023}, focal volume averaging was included in the calculation of nonadiabatic alignment dynamics of \ce{Na2}(\triplet). The results showed no significant difference compared to the results for a calculation without focal volume averaging.}. Finally, we normalize the calculated \cost$(t)$ traces such that the highest degree of alignment reached matches that observed in the experiment for the respective dimer. The purpose of this normalization is to ease visual comparison between experiment and theory. We note that the calculations are carried out assuming that the dimers reside at a position of the droplets where the local surface place is parallel to the detector plane, see the Appendix for details.

\subsection{Calculations for \ce{K2}}

\begin{figure*}
\includegraphics[width=17.8 cm]{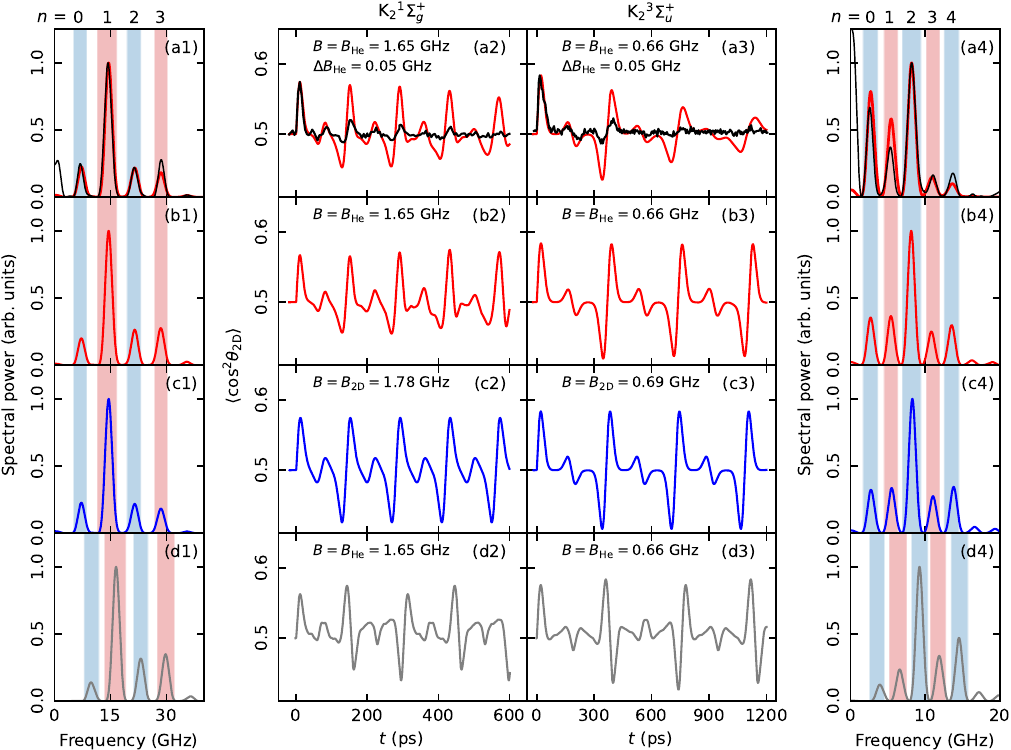}
\caption{(a2)--(a3) Experimentally measured \cost{} traces for \ce{K2} in the \singlet state and the \triplet state, shown in black. The red curves show the calculated results obtained with the 3D + \vhe model, \autoref{eqn:hamil}, including inhomogeneous broadening. (b2)--(b3) Simulation for the 3D + \vhe model, without inhomogeneous broadening. (c2)--(c3) Simulation for the isolated molecule with a 2D model, closely resembling the results of the 3D + \vhe model. (d2)--(d3) Simulation for the isolated molecule with a 3D model. (a1)--(d1), (a4)--(d4) Power spectra of the corresponding \cost{} traces in (a2)--(d2) and (a3)--(d3), respectively.  The spectral peaks are assigned numbers $n$, given above panel (a1) and (a4), in order and highlighted with colored bands (blue: even $n$, red: odd $n$).}
\label{fig:K2_traces}
\end{figure*}

The red curve in \autoref{fig:K2_traces}(a2) shows \cost$(t)$ for \ce{K2} in the \singlet state obtained for the parameters, \bhe~=~1.65~GHz and \deltaB~=~0.05~GHz, which led to the best agreement with the experimental \cost$(t)$ in terms of matching the positions of the revival structures. We note that \bgas~=~1.68~GHz~\cite{deiglmayr_calculations_2008}, so the in-helium rotational constant is only slightly reduced compared to the gas-phase value, \bgas/\bhe~=~1.02. The gradual decay of the calculated \cost$(t)$ results from the
inhomogeneous broadening applied as seen when comparing the calculation in \autoref{fig:K2_traces}(a2) with the one in \autoref{fig:K2_traces}(b2), where inhomogeneous broadening was omitted. Experimentally, the amplitude of \cost$(t)$ decreases significantly faster than for the calculated \cost$(t)$. Increasing \deltaB in the calculation leads to a faster decay of the amplitude but it also shifts the minima and maxima of the revivals in ways that cannot be brought to match the experimental \cost$(t)$ by adjusting the value of \bhe. Thus, changing \deltaB does not give an overall better agreement with the experiment. This indicates that homogeneous broadening also contributes to the gradual decrease of the degree of alignment observed in the experiment (see discussion in \autoref{sec:discussion}).

We also Fourier transformed the simulated \cost$(t)$ to enable a comparison with the experimental spectrum. The results are depicted in \autoref{fig:K2_traces}(a1) with the height of the tallest peak in each spectrum normalized to unity. We find that the central positions of the peaks and their relative amplitudes are very similar for the calculated and the measured spectrum.

\Autoref{fig:confinement}(b1) shows that the \vhe potential localizes the dimer axis to a region centered at the surface plane of the droplet. To test whether this localization is sufficiently strong that the laser-induced rotation effectively occurs in a plane, we solved the time-dependent Schrödinger equation for \ce{K2} in the \singlet state using an isolated 2D rotor model~\cite{mirahmadi_quantum_2021}. Again, we treat the effective rotational constant, now denoted \btd, as a free parameter. \Autoref{fig:K2_traces}(c2) displays the simulated alignment trace, obtained for the value of \btd, 1.78~GHz, that gives the best match with the experimental alignment trace. We interpret \btd as the effective rotational constant with respect to the normal vector of the surface, which is the relevant axis of rotation for the 2D motion. Since the dimer is not lying completely flat on the surface, according to \autoref{fig:confinement}(b1), we expect that \btd is larger than \bhe. In fact, from \ptheta we can determine the (classically) expected ratio \btd/\bhe as $\langle B\rangle / B$, where $\langle B\rangle$ is the value of the rotational constant $B$ averaged over \ptheta. From \ptheta in \autoref{fig:confinement}(b1), we find $\langle B\rangle / B=1.15$, which is fairly close to \btd/\bhe~=~1.08.

As a reference, we calculated the alignment dynamics of isolated gas-phase \ce{K2}(\singlet) dimers, that is, with \vhe removed from the Hamiltonian in \autoref{eqn:hamil}. The result obtained with a rotational constant $B$~=~\bhe, shown in \autoref{fig:K2_traces}(d2), looks qualitatively different from the measured \cost$(t)$. Notably the polarity of every second revival structure is opposite to that in the experimental trace. This difference persists independent of the choice of $B$. Similarly, the spectrum corresponding to the isolated gas-phase dynamics cannot be brought to match the experimental spectrum no matter the choice of $B$. Thus, the observed alignment dynamics cannot be described by the laser-induced rotation of isolated molecules.

Next, we turn to the results for \ce{K2} in the \triplet state. The alignment trace calculated with the Hamiltonian in \autoref{eqn:hamil}, using \vhe displayed in \autoref{fig:confinement}(b2), is illustrated in \autoref{fig:K2_traces}(a3). The calculated \cost$(t)$ was obtained for \bhe~=~0.66~GHz and \deltaB~=~0.05~GHz, which are the values that produced the best agreement with the experimental \cost$(t)$ in terms of matching the positions of the revival structures. In comparison, \bgas~=~0.79~GHz~\cite{bauer_accurate_2019}, so \bgas/\bhe~=~1.20. Thus, the triplet state dimer is relatively more He-dressed than the singlet state dimer. As for \ce{K2}(\singlet), increasing \deltaB results in a better agreement with the experimental alignment concerning the decay in \cost{} but deteriorates the agreement with respect to the positions of the revivals.

Like in the \ce{K2}(\singlet) case we also applied the 2D rotor model. \Autoref{fig:K2_traces}(c3) shows the resulting alignment trace obtained for \btd~=~0.69~GHz, which led to revival maxima and minima at positions closely matching those in the experimental trace. The alignment dynamics from the 2D model is also essentially identical to \cost$(t)$ from the 3D + \vhe model without the inhomogeneous broadening included, \autoref{fig:K2_traces}(b3). As for \ce{K2}(\singlet), we determined $\langle B\rangle / B=1.03$ from \ptheta, which is quite similar to \btd/\bhe~=~1.05. Note that for the \ce{K2}(\triplet) dimers \btd is (percent-wise) closer to \bhe than in the case of the \ce{K2}(\singlet) dimers. The reason is, we believe, that the triplet dimers are lying more flat on the surface than the singlet dimers -- see \autoref{fig:confinement}(b1) and \autoref{fig:confinement}(b2).

\subsection{Calculations for \ce{Na2}}

For \ce{Na2} in the \triplet state, the best agreement with the experimental \cost$(t)$ is obtained for \bhe~=~1.27~GHz and \deltaB~=~0.1~GHz in the 3D + \vhe calculation, \autoref{eqn:hamil}, with \vhe displayed in \autoref{fig:confinement}(a2). The simulated \cost$(t)$ is illustrated by the red curve in \autoref{fig:all_traces}(a3). The value of the gas-phase rotational constant is 1.65~GHz~\cite{bauer_accurate_2019}, which gives \bgas/\bhe~=~1.30. Thus, \ce{Na2}(\triplet) experiences some He-dressing similar to what was observed for \ce{K2}(\triplet). We also applied the 2D rotor model to \ce{Na2} in the triplet state and found excellent agreement with the structure of the experimental alignment trace using \btd~=~1.33~GHz \footnote{The simulated result is not shown here but can be seen in Fig. 2 of Ref.~\cite{kranabetter_nonadiabatic_2023}}. As for the \ce{K2} dimer, we determined $\langle B\rangle / B$ from \ptheta and found a ratio of 1.05, which is spot on with \btd/\bhe~=~1.05.

We also simulated the alignment dynamics for \ce{Na2} in the \singlet state with the 3D + \vhe model now using \vhe shown in \autoref{fig:confinement}(a1). Here, the best agreement with the experimental \cost$(t)$ comes with the parameters \bhe~=~4.53~GHz and \deltaB~=~0.17~GHz. For comparison, \bgas~=~4.64~GHz~\cite{deiglmayr_calculations_2008} so \bgas/\bhe~=~1.02, which follows the trend of only very little He-dressing found for \ce{K2}(\singlet). Interestingly, the 2D rotor model was not able to qualitatively capture the experimental alignment trace. We believe this is due to the much wider \ptheta of \ce{Na2}(\singlet) compared to the other dimers, see \autoref{fig:confinement}, i.e., the angular localization of the dimer axis to the surface plane is not strong enough to restrict the rotation to two dimensions.

\subsection{Calculations for \ce{Rb2}}

The red curve in \autoref{fig:all_traces}(c2) shows the calculated \cost$(t)$ for \ce{Rb2} in the \singlet state, based on the 3D + \vhe model using \vhe displayed in \autoref{fig:confinement}(c1). In this case, we find that it is \bhe~=~0.66~GHz and \deltaB~=~0.05~GHz that give the best match with the experimental alignment trace. The gas-phase rotational constant is 0.67~GHz~\cite{deiglmayr_calculations_2008} so \bgas/\bhe~=~1.02, in line with the findings for the singlet state of \ce{K2} and \ce{Na2}. We also find that the rotational dynamics are well-described by the 2D rotor model (not shown) for \btd~=~0.69~GHz. This is expected since the internuclear axis of \ce{Rb2}(\singlet) is quite strongly confined to the surface plane of the droplet, see \autoref{fig:confinement}(c1). Again, we determined $\langle B\rangle / B$ from \ptheta and found a ratio of 1.08, which is quite close to \btd/\bhe~=~1.05.

For \ce{Rb2} in the \triplet it is not possible to obtain a reliable match between the calculated and the measured \cost$(t)$ due to the lack of alignment dynamics after the initial alignment peak in the measurement, see \autoref{fig:all_traces}(c3). Instead we employ a previously calculated effective rotational constant for the dimer at a helium surface, \bhe~=~0.28~GHz~\cite{guillon_theoretical_2011}, a value which is slightly smaller than \bgas~=~0.32~GHz~\cite{bauer_accurate_2019}. Using this \bhe and \deltaB~=~0.05~GHz, the same amount of inhomogeneous broadening as found for \ce{Rb2}(\singlet)  and for \ce{K2}, we calculate an alignment trace with the 3D + \vhe model. The result is presented as the red curve in \autoref{fig:all_traces}(c3). The calculation shows that \cost{} remains at $\sim$~0.5 in the interval from $t$~=~100~ps to $t$~=~700~ps followed by a revival with moderate amplitude. Presumably, this revival structure is not observed in the measurement due to the same mechanisms that are responsible for the gradual decrease in the degree of alignment for the other alkali dimers. We note that alignment dynamics are still observed in the measurement for \ce{Rb2} in the \singlet state around the time scale of the missing half revival in the \triplet state measurement, as can be seen by comparing \autoref{fig:all_traces}(c2) and (c3). This indicates that the gradual decay in the degree of alignment happens faster for \ce{Rb2} in the \triplet state than for the \singlet state.

\section{Discussion}\label{sec:discussion}

\Autoref{tab:B-values} summarizes the effective \bhe values obtained for the different alkali dimers via fits to the 3D model including \vhe. To assess the uncertainties in the reported \bhe values, we have done a series of additional calculations with slight changes to \bhe. Based on how well the calculations fit the measured alignment traces, we estimate the uncertainty in \bhe to 0.02~GHz. If \bhe is varied more than that, we observe significant discrepancies between the calculations and the measured alignment traces, in particular for the central positions of the peaks in the spectra.

Via the experimentally determined \bhe values and the theoretical values for \bgas we have calculated the ratios \bgas/\bhe for each dimer, with the exception of \ce{Rb2} in the \triplet state. All ratios are presented in \autoref{tab:B-values}. The uncertainties indicated were derived via the estimated uncertainty of \bhe and do not include possible uncertainties of the literature values of \bgas. For the \singlet state dimers, \bgas is only 2\% larger than \bhe, indicating that the interaction with the helium droplet is very weak. In contrast, \bgas is 30\% and 20\% larger than \bhe for \ce{Na2} and \ce{K2} in the \triplet states, respectively, suggesting that the \triplet state dimers are more strongly influenced by the helium surface than their \singlet state counterparts. Finally, we note that the \bgas/\bhe ratios for both spin states are much smaller than the ratios of 2--6 observed for linear molecules inside helium droplets~\cite{choi_infrared_2006,chatterley_rotational_2020}.

The primary shortcoming of the 3D model with \vhe and inhomogeneous broadening is that the amplitude of the revivals in the calculated \cost$(t)$ does not decay fast enough compared to the experimental observations. As mentioned earlier, this indicates that the rotational wave packets are also influenced by homogeneous broadening, that is, by loss of population in the excited rotational levels due to the finite lifetimes, and possibly also by a direct loss of phase~\cite{choi_infrared_2006, blancafort-jorquera_rotational_2019}.
We note that molecules rotating inside helium droplets appears less influenced by inhomogeneous broadening~\cite{chatterley_rotational_2020}.
This difference is currently not clear, nor is it clear why \ce{Na2} appears to be subjected to more inhomogeneous broadening than \ce{K2} and \ce{Rb2}, especially for the \singlet state.

\begin{table}
\begin{ruledtabular}
\centering
\caption{Experimentally found \bhe values and \bgas values for the most abundant isotopologue of each dimer calculated using equilibrium distances \req from the literature. Also included are the ratios $B_{\text{gas}}$/$B_{\text{He}}$. The stated uncertainties in \bhe have been qualitatively estimated and propagated to \bgas/\bhe (see \autoref{sec:discussion}).}
\label{tab:B-values}
\begin{tabular}{@{}lccc}
		State 			 & \bhe (GHz) & \bgas (GHz) 		&	$B_{\text{gas}}$/$B_{\text{He}}$ \\ \midrule
\ce{Na_2} \singlet &     4.53 $\pm$ 0.02  		& 		4.64\footnotemark[1] 	&	1.02 $\pm$ 0.01\\
\ce{K_2}  \singlet &   	1.65 $\pm$ 0.02			& 		1.68\footnotemark[1] 	&	1.02 $\pm$ 0.01\\
\ce{Rb_2} \singlet &   	0.66 $\pm$ 0.02  		& 		0.67\footnotemark[1] 	&	1.02 $\pm$ 0.03\\	
\ce{Na_2} \triplet &  	1.27 $\pm$ 0.02  		& 		1.65\footnotemark[2] 	&	1.30 $\pm$ 0.02\\	
\ce{K_2}  \triplet &   	0.66 $\pm$ 0.02  		& 		0.79\footnotemark[2] 	&	1.20 $\pm$ 0.04\\	
\ce{Rb_2} \triplet &  	-		  				& 		0.32\footnotemark[2] &	-	\\	
\end{tabular}
\end{ruledtabular}
\footnotetext[1]{From Ref.~\onlinecite{deiglmayr_calculations_2008}.}
\footnotetext[2]{From Ref.~\onlinecite{bauer_accurate_2019}.}
\end{table}

\section{Conclusion and outlook}\label{sec:conclusion}

There are two major points, we want to emphasize from this work. Both bring about new insight, leave open questions and point towards new opportunities.
Firstly, we showed that the three alkali dimer species, \ce{Na2}, \ce{K2}, and \ce{Rb2}, residing on the surface of a helium nanodroplet and populated in either the \singlet or the \triplet state, can be set into rotation by a 1 ps laser pulse. In all three cases, the laser-dimer interaction leads to a global maximum in the degree of alignment at 5--25 ps after the laser pulse and with a peak value of \cost{} value in the range 0.57--0.66. At longer times, the alignment dynamics exhibit a number of transient structures (revivals) with a decreasing amplitude such that there is no discernible variation of \cost{} after 1000 ps. \ce{Rb2}(\triplet) makes an exception in that the only structure in \cost$(t)$ is the prompt peak; no revivals are observed. The simulated degree of alignment found by solving the time-dependent Schr\"odinger equation with the droplet-dimer potential included reproduces the experimental alignment dynamics well in terms of positions and polarity of the revival structures. Inclusion of inhomogeneous broadening, corresponding to a distribution of rotational constants, leads to a decreasing amplitude of the simulated alignment trace although not as pronounced as in the experimental data. Secondly, using the rotational coherence spectroscopy aspect of nonadiabatic laser-induced alignment, we determined the effective rotational constants \bhe for the dimers with the exception of \ce{Rb2}(\triplet). For \ce{Na2} and \ce{K2} in the \triplet state \bhe is reduced by 20-30~\% compared to \bgas while for \ce{Na2}, \ce{K2} and \ce{Rb2} in the \singlet state \bhe is only a couple of percent smaller than \bgas. Currently, there are no theoretical results to benchmark these results against.

We believe the current work opens several new possibilities. 1) Inclusion of rotational echo techniques~\cite{karras_orientation_2015,rosenberg_echo_2018,zhang_rotational_2019} may enable a distinction between homogeneous and inhomogeneous broadening, something which is not possible with the current measurements. Such an experimental extension could shed light on why the alignment transients disappear so quickly compared to that expected based on pure inhomogeneous broadening. 2) Rotational echo techniques typically involve two alignment pulses. Such a scheme also makes it possible to explore enhancement of the degree of alignment by optimizing the timing and intensity of the two laser pulses~\cite{leibscher_enhanced_2004,bisgaard_observation_2004,lee_two-pulse_2004}. The largest value of \cost{} observed was 0.66 in the case of \ce{Na2}(\triplet). This is a very modest degree of alignment. In terms of potential applications of the aligned molecules it would be highly beneficial with a significantly larger degree of alignment, which is something that alignment using two, or more, laser pulses may provide~\cite{cryan_field-free_2009,christiansen_alignment_2015}. 3) The \ce{Li2} dimer, which is formed in the \triplet state on He droplets under standard conditions~\cite{lackner_spectroscopy_2013}, is expected to be significantly less pinned to the droplet surface even compared to \ce{Na2}(\triplet). Preliminary studies in our group indicate that its alignment dynamics differs significantly from any of those observed here. 4) Finally, it should also be possible to explore nonadiabatic alignment of heteronuclear dimers of alkali atoms or of mixed alkali--alkaline earth atom dimers and use the results to determine effective rotational constants, none of which are currently known.

\begin{acknowledgments}
H.S. acknowledges support from The Villum Foundation through a Villum Investigator Grant No. 25886. We thank Jan Th{\o}gersen for expert help with the optics and the laser system.
\end{acknowledgments}

\appendix*\section{}
\label{sec:Appendix}

\section*{Details on theoretical modeling}
\subsection{3D model}

First, we note that the experimental \cost(t) is determined for those dimers that are parallel to, or close to parallel to, the detector plane. This is so because in the 2D velocity images, we only analyze \ce{Ak+} fragment ions in a fairly narrow ring around the maximum radius defined by the ions from Coulomb explosion of dimers aligned parallel to the detector plane, see \autoref{fig:images_potentials}(a) and (b), where the outermost ring (innermost) ring contains the \ce{K+} ions from the \singlet (\triplet) state.

In the simulations, we assume that the dimers are located at a position on the droplet where the surface plane is parallel to the detector plane. This emulates to some degree the outcome of the experiment and, in addition, it greatly simplifies the numerical calculations compared to if the surface plane is randomly oriented with respect to the detector. We are aware that there could be some contribution to the experimental signal from other locations of the dimers on the droplets. One implication of not including this contribution is that the calculations tend to overestimate $\cost$ but, importantly, the omission of other dimers locations does not change the rotational dynamics of the dimers, i.e. the spectral peaks are not influenced.

\begin{figure}
\centering
\includegraphics[width=8.6cm]{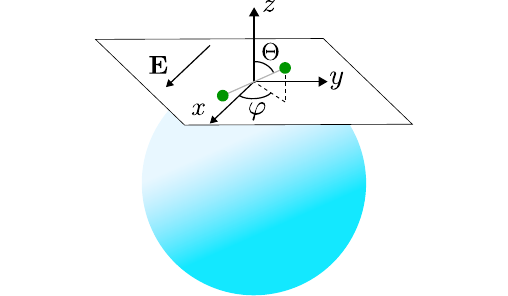}
\caption{Coordinate system used in the theoretical modeling. The dimer is represented by the two green spheres joined by the grey line, and the droplet is illustrated as a blue sphere (sizes are not to scale).}
\label{fig:droplet_sketch}
\end{figure}

The coordinate system used in the simulations is shown in \autoref{fig:droplet_sketch}. The detector plane is parallel to the $xy$ plane and the polarization of the alignment pulse is parallel to the x-axis, i.e. $\mathbf{E}\parallel \hat{\mathbf{x}}$. With this definition, the potential associated with the laser-molecule interaction can be written as \cite{sondergaard_nonadiabatic_2017}
\begin{equation}
V\left(\Theta, \varphi, t\right)=-\frac{E(t)^2}{4}\left(\Delta\alpha\sin^2\Theta\cos^2\varphi+\alpha_\perp\right),
\end{equation}
where $E(t)$ is the electric field amplitude and $\Delta\alpha=\alpha_\parallel-\alpha_\perp$ is the polarizability anisotropy. We employ a laser pulse with a Gaussian envelope function:
\begin{equation}
I(t)=I_0\exp\left[-4\ln{2}\left(\frac{t}{\tau}\right)^2\right],
\end{equation}
where $\tau$ is the full width at half maximum (FWHM) of the pulse duration and $I_0$ is the peak intensity. The full rotational Hamiltonian describing the alkali dimers on the droplet surface now takes the form
\begin{equation}
\hat{H}=\frac{B\hat{J}^2}{\hbar^2} + V_\text{He}(\Theta) + V\left(\Theta, \varphi, t\right) = \hat{H}_0 + V\left(\Theta, \varphi, t\right),
\end{equation}
which is equivalent to that given in \autoref{eqn:hamil}.

For a specific initial basis state of the dimer, the simulation proceeds by time propagating the state through diagonalization of the Hamiltonian with ($\hat{H}$) or without laser field ($\hat{H}_0$, for $|t|>1.5\tau$ the laser intensity is assumed to be zero) with a step size $dt=0.1\,\mathrm{ps}$ in the linear molecule basis $|JM\rangle$ (keeping up to $J_{\text{max}}=40$ states). Once the state is known at each time step, the degree of alignment observed in the detector plane can be obtained by calculating the expectation value of
\begin{equation}
    \cos^2\theta_\text{2D}=\cos^2\varphi.
\end{equation}
To take the temperature and nuclear spin statistics into account, we first diagonalize $\hat{H}_0$ to obtain the corresponding eigenvalues. We then do thermal averaging of $\langle \cos^2\theta_\text{2D}\rangle $ with the corresponding Boltzmann factor and include the abundances of even and odd $J$ states according to nuclear spin statistics (\vhe is even with respect to $\Theta$ around $\pi/2$ and does therefore not mix odd and even $J$ values). To account for the inhomogeneous broadening, the rotational constant is assumed to be normally distributed about $B_\text{0}$
\begin{equation}
p(B) = \frac{1}{\sqrt{2\pi}\sigma_B}\exp\left[-\frac{\left(B-B_0\right)^2}{2\sigma^2_B}\right],
\end{equation}
with a standard deviation $\sigma_B$, which is related to $\Delta B_\text{He}$ by $\Delta B_\text{He} = 2 \sqrt{2 \ln{2}} \sigma_B$. The expectation value $\langle \cos^2\theta_\text{2D}\rangle$ is averaged with respect to different rotational constants sampled from the normal distribution (200 samples. Finally, we also include the abundance ratios of different isotopologues. For the cases where the dimer is composed of different isotopes, for example \ce{^39K^41K}, there is no effect of nuclear spin statistics.

\subsection{2D model}
When the molecule is confined to the $xy$ plane, the motion becomes two-dimensional. In that case, the Hamiltonian of the system is
\begin{equation}
\hat{H}=B\hat{J}_z^2+ V\left(\varphi, t\right),
\end{equation}
with
\begin{equation}
V\left(\varphi, t\right)=-\frac{E(t)^2}{4}\left(\Delta\alpha\cos^2\varphi+\alpha_\perp\right).
\end{equation}
The definition of $\cos^2\theta_{\text{2D}}$ is the same as for the 3D case. For the 2D case, the basis states are now the eigenstates of $\hat{J}_z$, satisfying the relation $\hat{J}_z \ket{M}=\hbar M \ket{M}$ (again keeping the states from $M_\text{min}=-40$ to $M_\text{max}=40$), where now odd/even sector division is with respect to the quantum number $M$, since that is what defines parity in the 2D case.

%


\end{document}